\documentclass[aps, showkeys,showpacs,nofootinbib,floatfix]{revtex4}

\usepackage{amssymb}
\usepackage{amsfonts}
\usepackage{amsmath}
\usepackage{graphicx}
\usepackage{hyperref}
\usepackage{multirow,tabularx}

\begin{document}

\title{Fermions Analysis of IR modified Ho$\check{r}$ava-Lifshitz gravity:
Tunneling and Perturbation Perspectives}
\author{Molin Liu$^{1}$}
\thanks{Corresponding author\\E-mail address: mlliu@mail2.xytc.edu.cn}
\author{Junwang Lu$^{1}$}
\author{Jianbo Lu$^{2}$}
\affiliation{$^{1}$College of Physics and Electronic Engineering,
Xinyang Normal University, Xinyang, 464000, P. R. China\\
$^{2}$Department of Physics, Liaoning Normal University, Dalian, 116029, P. R. China}

\begin{abstract}
In this paper, we investigate the fermions Hawking radiation and quasinormal modes in infra-red modified Ho$\check{r}$ava-Lifshitz gravity under
tunneling and perturbation perspectives.
Firstly, through the fermions tunneling in IR modified Ho$\check{r}$ava-Lifshitz gravity,
we obtain the Hawking radiation emission rate, tunneling temperature and entropy
for the Kehagias-Sfetsos black hole. It is found that the results of fermions tunneling are in consistence with
the thermodynamics results obtained by calculating surface gravity.
Secondly, we numerically calculate
the lowing quasinormal modes frequencies of fermions perturbations by using WKB formulas including
the third orders and the sixth orders approximations simultaneously. It turns out that the actual frequency of
fermions perturbation is larger than in the Schwarzschild case, and the damping rate is smaller
 than for the pure Schwarzschild. The resluts of fermions perturbation suggest the quasinormal modes could be lived more longer in
 Ho$\check{r}$ava-Lifshitz gravity.
\end{abstract}



\maketitle

\section{Introduction}
Recently, Ho$\check{r}$ava presented a power counting renormalizable gravity theory at Lifshitz
point which is called
Ho$\check{r}$ava-Lifshitz (HL) gravity \cite{Horava11}. It exhibits a broken Lorentz symmetry at short distances
and reduces to usual general relativity (GR) gravity at the large distances with particular $\lambda = 1$
which controls the contribution of the extrinsic curvature trace. With HL gravity theory putting
forth, HL gravity is intensively investigated in many
aspects involving basic formalism \cite{basicformalism}, cosmology \cite{cosmology},
various black hole solutions and their thermodynamics \cite{BHsolutions,Kehagias,added1,added2} and so on.

In the subsequent developments of the HL gravity, people are trying to find the influence of matter fields for HL gravity. Which analogue of the matter energy-momentum tensor could be used to act gravitational source as GR? The pioneering works involve the geodesic analysis by various methods including the optical limit of a scalar field
theory \cite{Capasso}, super Hamiltonian formalism \cite{Rama},
foliation preserving diffeomorphisms \cite{Mosaffa},
Lorentz-violating Modified Dispersion Relations \cite{Sindoni} an so on. The optical limit presented by Capasso and Polychronakos \cite{Capasso} could offer a deformed geodesic equation by the generalized Klein-Gordon action. It is shown that the particles maybe does not move along geodesic. Deviations from geodesic motion appear both in flat and Schwarzschild-like spacetimes. Similar result deviations from GR also is
found in \cite{Rama,Mosaffa,Sindoni} with various above-mentioned
approaches.

Considering vanishing cosmological constant ($\Lambda_W$), Kehagias and Sfetsos \cite{Kehagias}
proposes an asymptotically flat black hole solution by introducing the addition term
proportional to the Ricci scalar of three geometry $\mu^4 R^{(3)}$, which indicates the Minkowski vacuum and
 modified GR at infra-red (IR) modification. Then after that, many people have devoted to
 its phenomenology involving strong field gravitational lensing\cite{Konoplya2,sbchen}, scalar
 field quasi-normal modes\cite{Konoplya2,sbchenqnm}, timelike geodesic motion \cite{jhchen},
 thin accretion disk\cite{Harko1} and observations constraints \cite{xianzhiguance} as well as
 its thermodynamics analysis \cite{Myung1,GUPjieshi1,castingjieshi2}, which presents the
black hole entropy $S = A/4 + \pi/\alpha\ln (A/4)$ via
the first law of thermodynamics where $\alpha$ is Ho$\check{r}$ava parameter.
This entropy could be treated as Generalized uncertainty principle quantum correction entropy \cite{GUPjieshi1}
 or casting entropy \cite{castingjieshi2}.
 If fermions are tunneling in this kind HL gravity, might it keep this logarithm entropy?
 Motivated by this, we investigate fermions tunneling in section III.


On the other hand, as we know that through some additional fields (e.g. scalar or fermions fields), the black hole suffers a damping
oscillation phase which is named as ``quasi-normal models"  (QNMs) or ``quasi-normal
ringing". As a results, the normal model oscillation is replaced by a complex frequencies which
encodes the black hole's important information such as mass, charge, momentum and the dimensions of
spacetime. The real part of complex frequency represents the actual frequency and the imaginary
part of its represents the damping of the oscillation. It is believed that these QN frequencies could be detected
by (LIGO, VIRGO, TAMT, GEO600) in future. So according to this observable QNMs, some people have used massless scalar field to obtain the important lowing QN frequencies which could live longer and be detected easily \cite{Konoplya2,sbchenqnm}. It is interesting that the QNMs of massless scalar field are longer lived and have larger real oscillation frequency in Ho$\check{r}$ava-Lifshitz gravity than in GR. Else, as one kind of basic particles, the fermions could offer us many important information. Motivated by the situations above, we will evaluate the QNMs for the massless fermions perturbations in section IV.

As we have already known that, due to the different kinetic terms, Ho$\check{r}$ava-Lifshitz gravity with $\lambda \neq 1$ is different significantly from General Relativity. Hence, the HL gravity with $\lambda \neq 1$ has been extensively studied in the literatures. The relevant works are mainly concentrated on the basic problems such as how to find various exact black hole solutions, the application of cosmology, the constraints of various fundamental parameters and so on. Despite more attention has been paid to the properties of HL black holes, there are a few works referring to the fermions analysis, especially to the $\lambda \neq 1$ case. So far as we know, the works relevant the fermion analysis focus mainly on the IR modified HL gravity including Dirac perturbations \cite{added31,added32} and fermion tunnelling for $z = 4$ black holes \cite{added33} and so on. In Wang and Gui's work \cite{added31}, the quasinormal frequencies of massless Dirac field
perturbation are evaluated by third-order WKB approximation. In Varghese and Kuriakose's work \cite{added32}, the evolution of Dirac perturbations is also investigated by using time domain integration and third-order WKB methods. In Chen, Yang and Zu's work \cite{added33}, the fermion tunnelling is investigated in the background of (3 + 1) dimensions and (4 + 1) dimensions black holes in the $z = 4$ HL gravity.

This paper is
organized as follows. In section II, we present the Kehagias and Sfetsos black hole solutions.
In section III, we calculate the fermions tunneling emission rate of Hawking radiation,
tunneling temperature and entropy. In section IV, we use the third and the sixth orders WKB
formulas to numerically calculate frequencies simultaneously. Section V is the conclusions.
We adopt the signature $(-, +, +, +)$ and put $\hbar$, $c$,
and $G$ equal to unity.
\section{An asymptotically flat infra-red modified black hole solution in deformed Ho$\check{r}$ava-Lifshitz gravity}
In this section, we review briefly the KS black hole solutions under
the limit of $\Lambda_{W} \longrightarrow 0$ with running constant
$\lambda = 1$ in the IR critical point $z = 1$. The space geometric
is parameterized with Arnowitt-Deser-Misner (ADM) formalism,
\begin{equation}\label{ADMmetric}
    d s^2 = - N^2 d t^2 + g_{ij} \left( d  x^i + N^i d t\right)\left(d x^j + N^j dt\right).
\end{equation}
The action for the fields of HL theory is

\begin{eqnarray}\label{action1}
\nonumber S &=& \int dt d^3 x \sqrt{g} N \bigg{\{} \frac{2}{\kappa^2} \left(K_{ij}K^{ij} - \lambda K^2 \right) - \frac{\kappa^2}{2 \alpha^4} C_{ij} C^{ij} + \frac{\kappa^2 \mu}{2 \alpha^2} \epsilon^{ijk} R_{il}^{(3)} \nabla_{j} R^{(3)l}_{\ \ \ \ k}\\
    && \ \ \ \ \ \ \ \ \ \ \ \ \ \ \ \ \ \ \ -\frac{\kappa^2 \mu^2}{8} R_{ij}^{(3)}R^{(3)ij} + \frac{\kappa^2 \mu^2}{8 \left(1 - 3\lambda\right)} \left(\frac{1 - 4 \lambda}{4}\left(R^{(3)}\right)^2 + \Lambda_{W} R^{(3)} - 3 \Lambda^2_{W}\right) + \mu^4 R^{(3)}\bigg{\}},
\end{eqnarray}
where the second fundamental form, extrinsic curvature $K_{ij}$, and
the Cotton tensor $C^{ij}$ are given as follows,
\begin{eqnarray}
  K_{ij} &=& \frac{1}{2N} \left(\frac{\partial }{\partial t}g_{ij} - \nabla_i N_j - \nabla_j N_i\right), \label{extrinsiccurvature}\\
  C^{ij} &=& \epsilon^{ikl} \nabla_k \left(R^{(3)j}_{l} - \frac{1}{4} R^{(3)} \delta_{l}^{j}\right). \label{Cottontensor}
\end{eqnarray}
Here, $\kappa$, $\lambda$, $\alpha$, $\mu$ and $\Lambda_{W}$ are the constant parameters. The last term of metric Eq.(\ref{action1}) represents a soft violation of the detailed balance
condition. Comparing the HL gravity action with that of GR gravity, we can obtain the speed of light $c$, the Newton's constant $G$ and the cosmological constant $\Lambda$
\begin{equation}\label{cGlambda}
    c = \frac{\kappa^2 \mu}{4} \sqrt{\frac{\Lambda_{W}}{1 - 3\lambda}},\ \ G = \frac{\kappa^2}{32\pi c},\ \ \Lambda = \frac{3}{2} \Lambda_{W}.
\end{equation}

In the limit of $\Lambda_{W} \longrightarrow 0$, we can obtain a deformed action as
follows,
\begin{eqnarray}
    S &=& \int dt_{L} d^3 x \left(\mathcal{L}_0 + \mathcal{L}_1\right),\label{deformedaction1}\\
    \mathcal{L}_0 &=& \sqrt{g} N \bigg{\{}\frac{2}{\kappa^2} \left(K_{ij}K^{ij} - \lambda K^2\right)\bigg{\}},\label{deformedaction2}\\
    \mathcal{L}_1 &=& \sqrt{g} N \bigg{\{} \frac{\kappa^2 \mu^2 \left(1 - 4\lambda\right)}{32\left(1 - 3\lambda\right)} \mathcal{R}^2 - \frac{\kappa^2}{2 \alpha^4}\left(C_{ij} -\frac{\mu \alpha^2}{2} R_{ij}\right) \left(C^{ij}-\frac{\mu \alpha^2}{2} R^{ij}\right) + \mu^4 \mathcal{R} \bigg{\}}.\label{deformedaction3}
\end{eqnarray}

For the particular case of $\lambda = 1$ with $\alpha = 16 \mu^2/\kappa^2$, a spherically symmetric black hole solution is presented by Kehagias and Sfetsos \cite{Kehagias}, which also is corresponding to an asymptotically flat space,
\begin{equation}\label{Kehagiassolution}
    d s^2 = -f(r) d t^2 + \frac{d r^2}{f(r)} + r^2 \left(d \theta^2 + \sin^2 \theta d \varphi^2 \right).
\end{equation}
The lapse function is
\begin{equation}\label{metricfunction}
    f(r) = 1 + \alpha r^2 - \sqrt{r\left(\alpha^2 r^3 +4 \alpha M\right)},
\end{equation}
where the parameter $M$ is an integration constant related with the mass of black hole. Using the null hypersurface condition, one can find there are two horizons, inner $r_-$ and outer event horizon $r_+$ in this space,
\begin{equation}\label{horizons}
    r_{\pm} = M \left(1 \pm \sqrt{1 - \frac{1}{2 \alpha M^2}}\right).
\end{equation}
Thermodynamic quantities including mass $M_{KS}$, temperature $T_{KS}$, and heat
capacity $C_{KS}$ and entropy $S_{KS}$ presented in Refs.\cite{Myung1} are listed as,
\begin{eqnarray}
  M_{KS} &=& \frac{1 + 2\alpha r_{\pm^2}}{4\alpha r_{\pm}}, \label{ADSmass}\\
  T_{KS} &=& \frac{2\alpha r_+^2 - 1}{8\pi r_+(\alpha r_+^2 + 1)}, \label{temperature}\\
  C_{KS} &=& -\frac{2\pi}{\alpha} \left[\frac{(\alpha r_+^2 + 1)^2 (2\alpha r_+^2 - 1)}{2\alpha^2 r_+^4 - 5\alpha r_+^2 - 1}\right],\label{heatcapacity}\\
  S_{KS} &=& \frac{A}{4} + \frac{\pi}{\alpha}\ln \left(\frac{A}{4}\right), \label{ksentropy}
\end{eqnarray}
with horizon area $A = 4 \pi r_+^2$. Under the limit of $\alpha \longrightarrow +\infty$, the entropy reduces
to Bekenstein-Hawking entropy $S_{BH} = A/4$ for Schwarzschild black hole.
\section{fermions tunneling of IR modified Ho$\check{r}$ava-Lifshitz gravity}

In this section, we investigate the Hawking radiation of Kehagias and Sfetsos
black hole in IR modified Ho$\check{r}$ava-Lifshitz gravity with fermion
tunneling. The tunneling probability, temperature and entropy are expected to be obtained.
The Dirac equation
 in the KS black hole spacetime can be written as
\begin{equation}\label{21diraceuation}
    \left[\gamma^{a} e_{a}^{\mu}\left(\partial_{\mu} + \Gamma_{\mu}\right) + \frac{m}{\hbar}\right] \Psi = 0,
\end{equation}
where $m$ is the mass of fermions. $e_{a}^{\mu}$ is the inverse of the tetrad $e^{a}_{\mu}$ defined by black hole metric
$g_{\mu\nu} = \eta_{ab}e^{a}_{\mu}e^b_\nu$ with Minkowski metric
$\eta_{ab} = \text{diag} (-1, 1, 1, 1)$. $\gamma^{a}$ is the Dirac matrix and
$\Gamma_{\mu}$ is the spin connection given by
\begin{equation}\label{21zixuanlianluo}
    \Gamma_{\mu} = \frac{1}{8}\left[\gamma^{a}, \gamma^{b}\right]e_{a}^{\nu}e_{b\nu ; \mu},
\end{equation}
where the covariant derivative of $e_{b\nu}$ is given by Christoffel symbols $\Gamma_{\mu\nu}^{a}$ as,
\begin{equation}\label{21weishang}
  e_{b\nu; \mu} =  \partial_{\mu} e_{b\nu} - \Gamma_{\mu\nu}^{a}e_{ba}.
\end{equation}
We choose following $\gamma$ matrix,
\begin{equation}\label{21ggggammametric}
\gamma^{0}=
\begin{pmatrix}
-i & 0 \\ 0 & i
\end{pmatrix},
\gamma^{1}=
\begin{pmatrix}
0 & -i\sigma^3 \\ i\sigma^3 & 0
\end{pmatrix},
\gamma^{2}=
\begin{pmatrix}
0 & -i\sigma^2 \\ i\sigma^2 & 0
\end{pmatrix},
\gamma^{3}=
\begin{pmatrix}
0 & -i\sigma^1 \\ i\sigma^1 & 0
\end{pmatrix},
\end{equation}
where $\sigma^{i}$ is Pauli sigma matrix,
\begin{equation}\label{21ppppaulimatrix}
\sigma^{1}=
\begin{pmatrix}
0 & \ \ 1 \\ 1 &\ \ 0
\end{pmatrix},
\sigma^{2}=
\begin{pmatrix}
0 & -i \\ i &\ \ 0
\end{pmatrix},
\sigma^{3}=
\begin{pmatrix}
1 &\ \ 0 \\ 0 & -1
\end{pmatrix}.
\end{equation}

In the presentation of $\sigma^{i}$, the spin up wave function is written as,
\begin{eqnarray}\label{21spinup}
 \nonumber   \psi_{\uparrow} (t,r,\theta,\phi) &=& \begin{pmatrix}
 A (t,r,\theta,\phi) \xi_{\uparrow}\\ B (t,r,\theta,\phi) \xi_{\downarrow}
\end{pmatrix} \exp \left[\frac{i}{\hbar} I_{\uparrow}(t,r,\theta,\phi)\right] \\
&=& \begin{pmatrix}
 A (t,r,\theta,\phi) \\ 0 \\ B (t,r,\theta,\phi) \\
 0 \end{pmatrix} \exp \left[\frac{i}{\hbar} I_{\uparrow}(t,r,\theta,\phi)\right],
\end{eqnarray}
where $\xi_{\uparrow}$ denotes the eigenvector of spin up state with eigenvalue $+1$,
and $\xi_{\downarrow}$ denotes the eigenvector of spin down state with eigenvalue $-1$.
$I_{\uparrow}$ is the action of radiation particles with spin up.
According to $e_{a}^{\nu}e_{\mu}^{a} = \delta_{\mu}^{\nu}$, we have
\begin{equation}\label{21eamu}
e_{a}^{\mu} = \text{diag} \left(\frac{1}{\sqrt{f}},\ \  \sqrt{f},\ \ \frac{1}{r},\ \ \frac{1}{r\sin\theta}\right).
\end{equation}
Submitting Eq.(\ref{21eamu}) into Dirac Eq.(\ref{21diraceuation}),
the frame $e_{a}^{\mu}$ should satisfy following relation
\begin{equation}\label{21biaojia}
    \left(\gamma^{0} e_{0}^{t} \partial_t + \gamma^{1} e_{1}^{r} \partial_r +
    \gamma^{2} e_{2}^{\theta} \partial_\theta +\gamma^{3} e_{3}^{\phi} \partial_\phi +
    \gamma^{a} e_{a}^{\mu} \partial_\mu \Gamma_{\mu} + \frac{m}{\hbar}\right) \Psi_{\uparrow} = 0.
\end{equation}
Simplifying above Eq.(\ref{21biaojia}), we can get
\begin{equation}\label{21simplifiedequa}
    \left(\frac{\gamma^0}{\sqrt{f}}\partial_t + \sqrt{f} \gamma^1 \partial_r +
    \frac{\gamma^2}{r} \partial_\theta + \frac{\gamma^3}{r\sin\theta} \partial_\phi +
    \gamma^a e_a^\mu \Gamma_\mu + \frac{m}{\hbar} \right) \Psi_{\uparrow} = 0.
\end{equation}
If we neglect the small quantity $\Gamma_{\mu}$, Eq.(\ref{21simplifiedequa}) could be simplified as,
\begin{equation}\label{21simpliequation}
    \left(\frac{\gamma^0}{\sqrt{f}}\partial_t + \sqrt{f} \gamma^1 \partial_r +
    \frac{\gamma^2}{r} \partial_\theta + \frac{\gamma^3}{r\sin\theta} \partial_\phi +
 \frac{m}{\hbar} \right) \Psi_{\uparrow} = 0.
\end{equation}
On the benefit of $\gamma$ metrics Eq.(\ref{21ppppaulimatrix}), we have
\begin{eqnarray}
\label{gamaphi1}  \gamma^0\partial_t \Psi_{\uparrow}  &=& \begin{pmatrix}
 A\frac{1}{\hbar} \exp\frac{iI_{\uparrow}}{\hbar} \partial_t I_{\uparrow}  \\
  0 \\
- B\frac{1}{\hbar} \exp\frac{iI_{\uparrow}}{\hbar} \partial_t I_{\uparrow}\\
 0 \end{pmatrix},\ \  \gamma^1 \partial_r \Psi_{\uparrow} = \begin{pmatrix}
 B\frac{1}{\hbar} \exp\frac{iI_{\uparrow}}{\hbar} \partial_r I_{\uparrow} \\
 0 \\
  - A\frac{1}{\hbar} \exp\frac{iI_{\uparrow}}{\hbar} \partial_r I_{\uparrow}\\
 0 \end{pmatrix},  \\
\label{gamaphi3} \gamma^2\partial_\theta \Psi_{\uparrow} &=& \begin{pmatrix}
0  \\  iB\frac{1}{\hbar} \exp\frac{iI_{\uparrow}}{\hbar} \partial_\theta I_{\uparrow} \\
0\\
-iA\frac{1}{\hbar} \exp\frac{iI_{\uparrow}}{\hbar} \partial_\theta I_{\uparrow} \end{pmatrix},\ \  \gamma^3 \partial_\phi \Psi_{\uparrow} = \begin{pmatrix}
0  \\
B\frac{1}{\hbar} \exp\frac{iI_{\uparrow}}{\hbar} \partial_\phi I_{\uparrow} \\
0\\
-A\frac{1}{\hbar} \exp\frac{iI_{\uparrow}}{\hbar} \partial_\phi I_{\uparrow} \end{pmatrix}.
\end{eqnarray}
Submitting Eqs.(\ref{gamaphi1}) and (\ref{gamaphi3}) into Dirac Eq.(\ref{21simpliequation}), we can get
\begin{equation}\label{21diaraccc}
    \frac{1}{\sqrt{f}}
    \begin{pmatrix}
A\partial_t I_{\uparrow}  \\
0 \\
-B\partial_t I_{\uparrow}\\
0 \end{pmatrix}
+ \sqrt{f}
\begin{pmatrix}
B\partial_r I_{\uparrow}  \\
0 \\
-A\partial_r I_{\uparrow}\\
0 \end{pmatrix}
+\frac{1}{r}
\begin{pmatrix}
0  \\
iB\partial_\theta I_{\uparrow} \\
0\\
-iA\partial_\theta I_{\uparrow}
\end{pmatrix}
+\frac{1}{r\sin\theta}
\begin{pmatrix}
0  \\
B\partial_\phi I_{\uparrow} \\
0\\
-A\partial_\phi I_{\uparrow}
\end{pmatrix}
+ m
\begin{pmatrix}
A \\
0 \\
B \\
0
\end{pmatrix}.
\end{equation}
This equation could be reduced to four components of $(t, r, \theta, \phi)$ as,
\begin{eqnarray}
  \frac{A}{\sqrt{f(r)}} \partial_{t} I_{\uparrow} + B \sqrt{f(r)} \partial_{r} I_{\uparrow} + m A&=& 0, \label{21action555}\\
  \frac{-B}{\sqrt{f(r)}} \partial_{t} I_{\uparrow} - A \sqrt{f(r)} \partial_{r} I_{\uparrow} + m B &=& 0, \label{21action666}\\
  \frac{B}{r} \left(i \partial_{\theta} I_{\uparrow}+ \frac{1}{\sin\theta}\partial_{\phi} I_{\uparrow}\right)&=& 0, \label{21action777}\\
  \frac{A}{r} \left(i \partial_{\theta} I_{\uparrow}+ \frac{1}{\sin\theta}\partial_{\phi} I_{\uparrow}\right)&=& 0.  \label{21action888}
\end{eqnarray}
Consider the symmetry of the spacetime, we adopt the action below as,
\begin{equation}\label{21action}
    I_{\uparrow} = - \omega t + \mathcal{W}(r) + \Theta (\theta, \phi).
\end{equation}
Submitting Eq.(\ref{21action}) into Eqs.(\ref{21action555}), (\ref{21action666}), (\ref{21action777}), (\ref{21action888}), we can get
\begin{eqnarray}
  -\frac{A}{\sqrt{f(r)}} \omega + B\sqrt{f(r)} \partial_{r} \mathcal{W} + m A &=& 0, \label{21aa555}\\
  \frac{B}{\sqrt{f(r)}} \omega - A\sqrt{f(r)} \partial_{r} \mathcal{W} + m B &=& 0, \label{21aa666}\\
  B \left(i \partial_{\theta} \Theta + \frac{1}{\sin\theta}\partial_{\phi} \Theta \right) &=& 0, \label{21aa777} \\
  A \left(i \partial_{\theta} \Theta + \frac{1}{\sin\theta}\partial_{\phi} \Theta \right) &=& 0.\label{21aa888}
\end{eqnarray}
Because the contribution of $\Theta$ on
outgoing particle is equal with that of $\Theta$ on incoming particles, Eqs.(\ref{21aa777}) and
(\ref{21aa888}) do absolutely nothing that are useful to the calculation of the tunneling probability,  We only need consider the action of the
radial direction, i.e. Eqs.(\ref{21aa555}) and (\ref{21aa666}), whose solvability condition is
the determinant of the coefficients of A and B is zero. Namely,
\begin{equation}\label{21hanglieshi}
    \begin{vmatrix}
-\omega\frac{1}{\sqrt{f}} + m & \ \ \sqrt{f} \partial_r \mathcal{W}\\
-\sqrt{f(r)}\partial_r \mathcal{W}\ \ & \omega\frac{1}{\sqrt{f}} + m \\
\end{vmatrix}
=0.
\end{equation}
By direct integration of determinant Eq.(\ref{21hanglieshi}), $\mathcal{W}$ could be obtained as,
\begin{equation}\label{21mathcalwww}
  \mathcal{W}_{\pm} (r) = \pm \int \frac{\sqrt{\omega^2 - m^2 f(r)}}{f(r)} dr.
\end{equation}
Using the condition $f(r)\longrightarrow 0$ near horizon $r_+$, the numerator of integrated
fraction in Eq.(\ref{21mathcalwww}) is reduced to $\sqrt{\omega^2 - m^2 f(r)}\longrightarrow \omega$.
Hence, the terms contained mass ($\sim m^2 f$) has nothing to do with the tunneling probability.
So, $\mathcal{W}_{\pm} (r)$ is applicable to the whole
fermions, no matter massive or massless particles. Adopting the contour integration, we can get
\begin{equation}\label{21integration}
 \mathcal{W}_{\pm} (r) =  \pm i \pi \frac{\omega}{f^{'}(r_+)},
\end{equation}
where $``+"$ denotes outgoing fermions, $``-"$ denotes incoming ones, $'$ means the
first-order derivation of $f(r)$ with respect to $r$,
\begin{equation}\label{21metricfunction}
f^{'}(r_+) = \frac{d f(r)}{d r} = 2\alpha r - \frac{2\alpha^2 r^3 + 2 \alpha M}{\sqrt{\alpha^2 r^4 + 4 \alpha M r}}.
\end{equation}

It is well known that the tunneling probability could be related to the imaginary
part of the action. Thus, the tunneling probability of the emission fermion is written as
followings,
\begin{equation}\label{21tunnellingprobability}
    \Gamma = \frac{P (emission)}{P (absorption)} = \frac{\exp (-2Im I_{\uparrow +})}{\exp (-2Im I_{\uparrow -})} = \frac{\exp (-2 Im \mathcal{W}_{+})}{\exp (-2 Im \mathcal{W}_{-})}.
\end{equation}
Submitting $\mathcal{W}_{\pm} (r)$ into above Eq.(\ref{21tunnellingprobability}), we can obtain
\begin{equation}\label{21gammmaaa}
\Gamma = \exp \left[-\frac{4 \pi \omega}{f^{'}(r_+)}\right] = \frac{2\pi r_+ \omega \left[\alpha r_+^3 + 4 M + \left(\alpha r_+^3 + M \alpha\right)\sqrt{1+ 4M/\alpha r_+^3}\right]}{4\alpha M r_+^2 - 2 \alpha^2 r_+^2 M - \alpha^2 M^2}.
\end{equation}
According to the usual relation between inverse temperature $\beta$ and tunneling probability,
 $\Gamma = \exp (- \beta \omega)$, we can get the fermions tunneling temperature,
\begin{equation}\label{21hawkingtemperture}
    T_{fermion} = \frac{f^{'}(r_+)}{4\pi} = \frac{1}{2\pi} \left[\alpha r_+ - \frac{\alpha^2 r_+^3 + \alpha M}{\sqrt{\alpha^2 r_+^4 + 4 \alpha M r_+}}\right].
\end{equation}
If we choose the mass function definded by Eq.(\ref{ADSmass}) in Ref.\cite{Myung1}, we can get
\begin{equation}\label{21temperature111}
    T_{fermion} = \frac{2\alpha r_+^2 - 1}{8\pi r_+ (1 + \alpha r_+^2)},
\end{equation}
which is just the Hawking temperature of the black hole in IR deformed Horava-Lifshitz
gravity \cite{Myung1}.
As a thermodynamical system, the first law $d M = T dS$ gives black hole entropy,
\begin{eqnarray}\label{211thentropy}
    S &=& \int T^{-1} d M + S_0 = \int dr_+ \left(\frac{1}{T}\frac{d M}{d r_+}\right) + S_0.\\
    &=& \pi \left(r_+^2 + \frac{1}{\alpha} \ln r_+^2\right) + S_0,
\end{eqnarray}
where we adopt $M_{KS}$ Eq.(\ref{ADSmass}). If we adopt $S_0 = \pi \ln \pi/\alpha$,
the final logarithmic entropy obtained through fermions tunneling is
\begin{equation}\label{211logentropy}
    S = \frac{A}{4} + \frac{\pi}{\alpha}\ln \frac{A}{4}.
\end{equation}

Based on the surface gravity defined by
\begin{equation}\label{surfacegravity}
 \kappa_+ = \frac{1}{2} \frac{d f }{d r}\bigg|_{r_{+}} = \frac{2\alpha r_+^2 - 1}{4r_+(1 + \alpha r_+^2)},
\end{equation}
the thermodynamic temperature Eq.(\ref{temperature})
and the thermodynamic entropy Eq.(\ref{ksentropy}) are
obtained in previous researches \cite{Myung1}. It is interesting that the resluts
Eqs.(\ref{temperature}) and (\ref{ksentropy}) based on surface gravity
are agreement with fermions tunneling results Eqs.(\ref{21temperature111}) and
(\ref{211logentropy}).
\section{fermions perturbations of IR modified Ho$\check{r}$ava-Lifshitz gravity}
In this section, we evaluate the quasinormal modes of fermions perturbation by using the third-order
and sixth-order WKB formulas, simultaneity. In order to get the quasinormal frequencies, we
should proceed from the Dirac Eq.(\ref{21diraceuation}).
According to the relation of $e_{b\mu} = \eta_{ab}e_{\mu}^{a}$, we can have
$e_{b\mu} = \text{diag} (-\sqrt{f}, 1/\sqrt{f}, r, r\sin\theta)$.
Then, based on
$e_{b\nu;\mu} = \partial_\mu\partial_{b\nu} - \Gamma_{\mu\nu}^{\alpha} e_{b\alpha}$,
the nonzero covariant derivative could be listed as,
\begin{eqnarray}
\nonumber  e_{01;0} &=& \frac{\sqrt{f}}{2f} f',\ \ \ \ e_{10;0} = \frac{-ff'}{2\sqrt{f}}, \ \ \ \ e_{21;2} = -1,\ \ \ \ e_{12;2} = \sqrt{f} r,\\
\label{22zixuanweishang2}  e_{23;3} &=& r\sin\theta\cos\theta,\ \ \ \ e_{13;3} = \sqrt{f} r \sin^2\theta,\ \ \ \ e_{31;3} = -\sin\theta,\ \ \ \ e_{32;3} = -r\cos\theta.
\end{eqnarray}
Submitting above Eqs.(\ref{22zixuanweishang2}) into
Eq.(\ref{21zixuanlianluo}), we can get the spin connections as,
\begin{equation}\label{22zixuanlianluo2}
    \Gamma_{t} = -\frac{1}{4} f' \gamma^0 \gamma^1;\ \ \Gamma_r = 0;\ \ \Gamma_{\theta} = -\frac{\sqrt{f}}{2}\gamma^1\gamma^2;\ \  \Gamma_\phi = -\frac{1}{2}\left(\sin\theta\sqrt{f}\gamma^1\gamma^3 + \cos\theta\gamma^2\gamma^3 \right).
\end{equation}
Considering the symmetry of frame Eq.(\ref{21eamu}), the Dirac Eq.(\ref{21diraceuation}) could be
 rewritten as,
\begin{equation}\label{22diracequation}
    \left[\gamma^0e_0^t \left(\partial_t + \Gamma_t\right)
    + \gamma^1e_1^r\left(\partial_r + \Gamma_r\right) + \gamma^2e_2^\theta\left(\partial_\theta +
    \Gamma_\theta\right) + \gamma^3e_3^\phi\left(\partial_\phi + \Gamma_\phi\right)\right]\Phi=0,
\end{equation}
where we adopt the massless Dirac
field to simplify the perturbation problem.

Based on the spin connections Eq.(\ref{22zixuanlianluo2}) and the anticommutation
relation of $\gamma$ metrics, Eq.(\ref{22diracequation}) could be reduced to a simple form
as following,
\begin{equation}\label{22diracequation22}
    \frac{\gamma^0}{\sqrt{f}}\frac{\partial \Phi}{\partial t} +
    \sqrt{f}\gamma^1 \left(\frac{\partial}{\partial r} +
    \frac{1}{r}\right)\Phi+\frac{\gamma^2}{r} \left(\frac{\partial}{\partial\theta} +
    \frac{1}{2}\cot\theta\right)\Phi + \frac{\gamma^3}{r\sin\theta}\frac{\partial \Phi}{\partial \phi} = 0,
\end{equation}
where $\Phi (t,r,\theta,\phi)= f^{-1/4}(r)\Psi(t,r,\theta,\phi)$.
Then, we could adopt an ansatz as followings,
\begin{equation}\label{22Phitihuan}
    \Phi = \frac{\Omega\left(\theta,\phi\right)}{r\sqrt{\sin\theta}}e^{-i\omega t}\begin{pmatrix}
F(r) \\
F(r) \\
iG(r) \\
iG(r)
\end{pmatrix}.
\end{equation}

\begin{figure}
  \includegraphics[width=4.5 in]{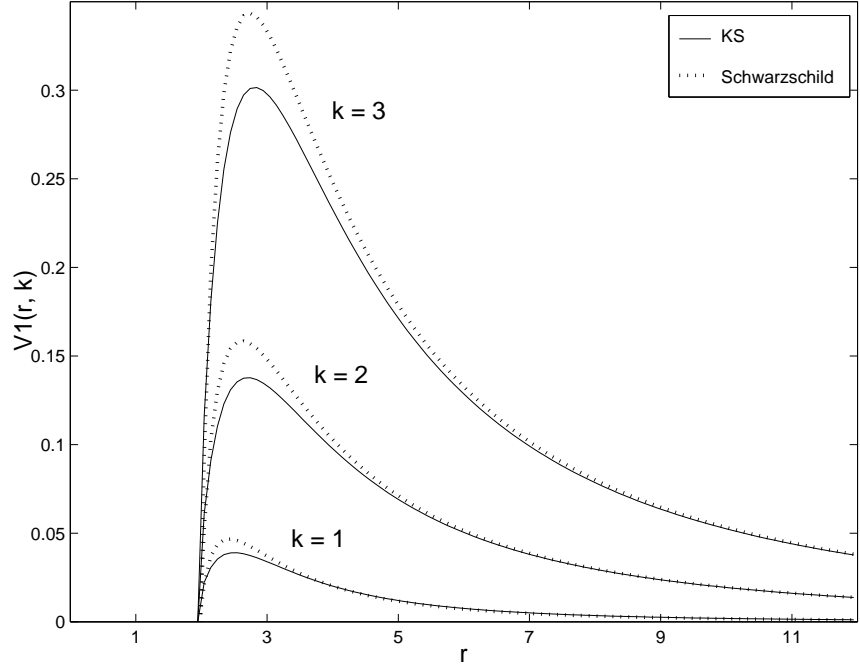}\\
  \caption{the potentials $V_1(r, k)$ (solid line) versus radial coordinate r with $k = 1, 2, 3$.
Meanwhile, we also draw Schwarzschild case with dotted line for comparison.}\label{potential}
\end{figure}

Submitting the ansatz Eq.(\ref{22Phitihuan}) into Eq.(\ref{22diracequation22}),
we can get three equations: one equation of $\Omega$ refers to variables $(\theta, \phi)$
and two equations of
$G(r)$ and $F(r)$ refer to variable $r$, which are listed as,
\begin{eqnarray}
 \label{22thetaphi222} -i\gamma^1 \gamma^0 \left(\gamma^2\partial_\theta + \frac{\gamma^3}{\sin\theta}\right)\Omega (\theta, \phi)
     &=& k \Omega (\theta, \phi),\\
 \label{22fg1} -\omega F(r) + \frac{d G(r)}{d r_{*}} - \frac{k\sqrt{f(r)}}{r} G(r) &=& 0, \\
 \label{22fg2}  \omega G(r) +  \frac{d F(r)}{d r_{*}} + \frac{k\sqrt{f(r)}}{r} F(r) &=& 0,
\end{eqnarray}
where $k = -l$ or $l+1$ and the coordinate transformation $d r = f (r) d r_{*}$ is adopted.
Eliminating $F(r)$ (or $G(r)$) in Eqs.(\ref{22fg1}) and (\ref{22fg2}), we can obtain two 2th
order differential equations of $G(r)$ (or $F(r)$),
\begin{eqnarray}
  \frac{d^2 F}{d r_{*}^2} + \left(\omega^2 - V_1\right) F &=& 0, \label{22fgfg11} \\
  \frac{d^2 G}{d r_{*}^2} + \left(\omega^2 - V_2\right) G &=& 0, \label{22fgfg22}
\end{eqnarray}
where $V_1$ and $V_2$ are supersymmetric partners with same spectra,
\begin{eqnarray}
\label{potential1}  V_1 &=& \frac{\sqrt{f(r)}|k|}{r^2} \left(|k|\sqrt{f(r)} + \frac{r}{2}\frac{d f(r)}{d r} - f(r)\right) \ \ k = l + 1,\\
\label{potential2}  V_2 &=& \frac{\sqrt{f(r)}|k|}{r^2} \left(|k|\sqrt{f(r)} - \frac{r}{2}\frac{d f(r)}{d r} + f(r)\right) \ \ k = -l.
\end{eqnarray}

In the following words, we use the Eq.(\ref{22fgfg11}) contained potential $V_1$ to evaluate the
quasinormal mode frequencies of the massless Dirac field by the third orders and sixth orders WKB approximation.
Here, $V_1 (r, k)$ is plotted in Fig.\ref{potential} which illustrates clearly that with bigger
$|k|$, $V_1$ is higher than that of Schwarzschild case (dotted lines). Moreover, the gap between
KS and Schwarzschild increases greatly as increasing $k$, in particular
near the maximum points. So we can expect the QNMs of fermions perturbations could be lived more longer and
the actual frequencies increase because there are more lower potential for IR modified Ho$\check{r}$ava-Lifshitz gravity.

According to the potential $V_1 (r, k)$ Eq.(\ref{potential1}),
the massless Dirac quasinormal modes in the KS
black hole spacetime satisfies the boundary conditions,
\begin{equation}\label{22qnmcondition}
    \Phi (x) \sim \exp (\pm i \omega ), \ \ \ \ x\longrightarrow\pm\infty.
\end{equation}
where $\omega = \text{Re} (\omega) + i \text{Im} (\omega)$. The real part
$\text{Re} (\omega)$ determines its actual oscillation frequency and the absolute
value of imaginary part $|\text{Im} (\omega)|$ determines the damping rate.

\begin{table}[!h]\label{table1}
\caption{Various low-lying overtones QN frequencies for a fixed $\alpha = 0.5$.}
\begin{tabular*}{0.78\textwidth}{ccccc}
     \hline
     \hline
    ~~~~$|k|$~~~~& ~~~~$n$~~~~&~~~~~~~~~~~~~~~$3th$~~~~~~~~~~~~~~~&
    ~~~~~~~~~~~~~~~$6th$~~~~~~~~~~~~~~~&~~~~~~$\text{Schwarzschild(6th)}$~~~~~~\\
\hline
     $1$&$0$&$0.200527 -0.071134 i$ &$0.199480 - 0.065511 i$&$0.182642 - 0.094937 i$\\
     $2$&$0$&$0.419782 -0.070205 i$&$0.420578 -0.069468 i$&$0.380069 - 0.096366 i$\\
     &$1$&$0.393067 - 0.213644 i$&$0.396805 - 0.208570 i$&$0.355860 - 0.297269 i$\\
    $3$&$0$&$0.635027 - 0.070293 i$&$0.635293 - 0.070174 i$&$0.574094 - 0.096307 i$\\
     &$1$&$0.617529 - 0.211964 i$&$0.618963 - 0.211069 i$&$0.557016 - 0.292717 i$\\
     &$2$&$0.583072 - 0.356887 i$&$0.586030 - 0.353696 i$&$0.526534 - 0.499713 i$\\
     $4$&$0$&$0.849148 - 0.070367 i$&$0.849265 - 0.070339 i$&$0.767354 - 0.096270 i$\\
     &$1$&$0.836227 - 0.211639 i$&$0.836870 - 0.211407 i$&$0.754300 - 0.290969 i$\\
     &$2$&$0.810528 - 0.354547 i$&$0.811997 - 0.353679 i$&$0.729754 - 0.491909 i$\\
     &$3$&$0.772412 - 0.500217 i$&$0.774470 - 0.498069 i$&$0.696728 - 0.702337 i$\\
     $5$&$0$&$1.062830 -0.070410 i$&$1.062890 -0.070401 i$&$0.960293 - 0.096254 i$\\
     &$1$&$1.052590 -0.211548 i$&$1.052930 -0.211469 i$&$0.949759 - 0.290149 i$\\
     &$2$&$1.032170 -0.353663 i$&$1.032970 -0.353352 i$&$0.929491 - 0.488114 i$\\
     &$3$&$1.001690 -0.497438 i$&$1.002930 -0.496649 i$&$0.901073 - 0.692514 i$\\
     &$4$&$0.961423 - 0.643574 i$&$0.962684 - 0.642041 i$&$0.866730 - 0.905116 i$\\
     \hline
     \hline
\end{tabular*}
\end{table}

In the various methods to get the frequcies of QNMs, the WKB numerical formulas are convenient to give
accurate frequencies values for the longer lived quasinormal models. This method is originally shown by
Schutz et al \cite{Schutz} and is later developed to the third order by Iyer et al
\cite{Iyer1,Iyer2}. At a later time, WKB approximation of QNMs is explanded to the sixth
order by Konoplya \cite{Konoplya1}. Then after that, this method
is extensively used in various spacetimes \cite{wkbyingyong}. In this paper, we numerically
calculate
the lowing modes frequencies through the sixth order WKB formula which has the
form \cite{Konoplya1} as following,
 \begin{equation}\label{6thorderwkb}
    \frac{i Q_0}{\sqrt{2Q_0^{''}}} - \Lambda_2 - \Lambda_3 - \Lambda_4 -\Lambda_5 - \Lambda_6 = n + 1/2,
 \end{equation}
where $Q$ is a ``reverse potential" given by $Q = \omega^2 - V$. $Q_0^i$ denotes the i-th derivative of $Q$ at its maximum point
with respect to the ``tortoise coordinate" $r_*$. The results of the third orders could also be obtained
by Eq.(\ref{6thorderwkb}) without $\Lambda_4$, $\Lambda_5$ and $\Lambda_6$. Considering WKB approximation fails
to calculate the higher order modes, we only evaluate low-lying QNM modes ($n < k$) by various overtones $n$.
The correctional terms of $\Lambda_2$ and $\Lambda_3$ are
given in Refs.\cite{Iyer1,Iyer2}. The correctional terms of $\Lambda_4$, $\Lambda_5$ and
$\Lambda_6$ are given in Ref.\cite{Konoplya1}. It turns out that WKB series shows well convergence
in all sixth orders for Dirac field, which is similar to the scalar field case
\cite{Konoplya2,sbchenqnm}. In this paper, we analyse
the effect of parameter Horava-Lifshitz gravities on QNM modes through two kinds of $\alpha$:
 one is fixed and another
is changed.

\begin{figure}
  \includegraphics[width=4.5 in]{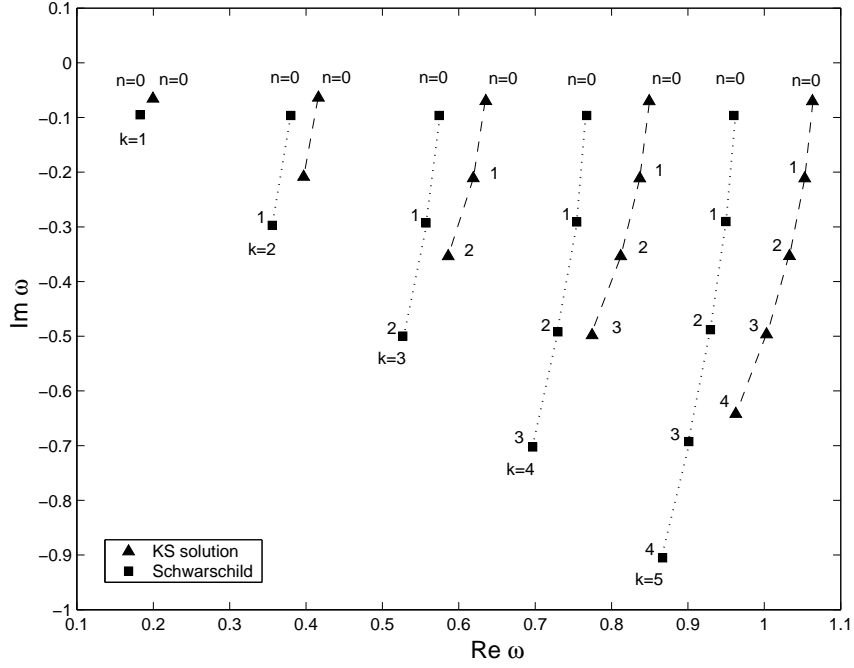}\\
  \caption{Massless Dirac quasinormal mode frequencies. For the convenience of comparison,
  double results including KS solution (solid triangle) and Schwarzschild solution
  (solid square) are given simultaneously.}\label{fig1}
\end{figure}
\begin{table}[!h]\label{22}
\caption{The QN frequencies of null overtones modes ($n = 0, k = 1$).}
\begin{tabular*}{0.5\textwidth}{ccc}
     \hline
     \hline
     ~~~~~$1/2\alpha$~~~~~ & ~~~~~~~~~~~~~~~$3th$~~~~~~~~~~~~~~~ & ~~~~~~~~~~~~~~~$6th$~~~~~~~~~~~~~~~\\
     \hline
     $0$ &$0.176452 - 0.100109 i$&$0.182642 - 0.094937 i$ \\
    $0.1$&$0.179189 - 0.098175 i$&$0.184899 - 0.092152 i$\\
    $0.2$&$0.181871 - 0.096140 i$&$0.187179 - 0.089352 i$\\
    $0.4$&$0.187091 - 0.091667 i$&$0.191448 - 0.083907 i$\\
    $0.5$&$0.189632 - 0.089156 i$&$0.193342 - 0.081265 i$\\
    $0.6$&$0.192118 - 0.086395 i$&$0.195055 - 0.078615 i$\\
    $0.8$&$0.196809 - 0.079810 i$&$0.197915 - 0.072858 i$\\
      $1$&$0.200527 - 0.071134 i$&$0.199480 - 0.065511 i$\\
     \hline
     \hline
\end{tabular*}
\end{table}
\begin{table}[!h]\label{33}
\caption{The QN frequencies of null overtones modes ($n = 0, k = 2$).}
\begin{tabular*}{0.5\textwidth}{ccc}
     \hline
     \hline
     ~~~~~$1/2\alpha$~~~~~ & ~~~~~~~~~~~~~~~$3th$~~~~~~~~~~~~~~~ & ~~~~~~~~~~~~~~~$6th$~~~~~~~~~~~~~~~\\
     \hline
     $0$ &$0.378627 - 0.0965424 i$&$0.380069 - 0.096366 i$ \\
    $0.1$&$0.381664 - 0.0949046 i$&$0.383040 - 0.094794 i$\\
    $0.2$&$0.384885 - 0.0931728 i$&$0.386218 - 0.093061 i$\\
    $0.4$&$0.391946 - 0.0893179 i$&$0.393230 - 0.089069 i$\\
    $0.5$&$0.395827 - 0.0871212 i$&$0.397087 - 0.086766 i$\\
    $0.6$&$0.399977 - 0.0846759 i$&$0.401200 - 0.084209 i$\\
    $0.8$&$0.409203 - 0.0786964 i$&$0.410286 - 0.078038 i$\\
      $1$&$0.419782 - 0.0702051 i$&$0.420578 - 0.069468 i$\\
     \hline
     \hline
\end{tabular*}
\end{table}
\begin{table}[!h]\label{44}
\caption{The QN frequencies of null overtones modes ($n = 0, k =3$).}
\begin{tabular*}{0.5\textwidth}{ccc}
     \hline
     \hline
     ~~~~~$1/2\alpha$~~~~~ & ~~~~~~~~~~~~~~~$3th$~~~~~~~~~~~~~~~ & ~~~~~~~~~~~~~~~$6th$~~~~~~~~~~~~~~~\\
     \hline
     $0$ &$0.573685 - 0.096324  i$&$0.574094 - 0.0963070  i$\\
    $0.1$&$0.578089 - 0.094773  i$&$0.578503 - 0.0947834  i$\\
    $0.2$&$0.582754 - 0.093108  i$&$0.583169 - 0.0931358  i$\\
    $0.4$&$0.592999 - 0.089343  i$&$0.593402 - 0.0893718  i$\\
    $0.5$&$0.598659 - 0.087176  i$&$0.599051 - 0.0871898  i$\\
    $0.6$&$0.604747 - 0.084754  i$&$0.605126 - 0.0847448  i$\\
    $0.8$&$0.618500 - 0.078810  i$&$0.618839 - 0.0787424  i$\\
      $1$&$0.635027 - 0.070293  i$&$0.635293 - 0.0701738  i$\\
     \hline
     \hline
\end{tabular*}
\end{table}

For the fixed $\alpha$, we only consider the first five low-lying modes $k = 1,\ 2,\ 3,\ 4,\ 5$ with
 $0 \leq n < k$. The results are listed in Table I. The quasinormal mode frequencies
for positive $k$ are plotted in Fig.\ref{fig1} which illustrates the real part
$Re \omega$ decreases with increasing mode
number $n$ for the given angular momentum number
$k$. Else, the absolute value of imaginary part $|Im \omega|$ increases as bigger $n$
which indicates
higher modes decay faster than the low-lying ones. Comparing with Schwarzschild results
(solid quare points), the $Re \omega$ of $\alpha$ is larger than Schwarzschild
limit, while the damping rate $|Im \omega|$ is smaller than pure Schwarzschild case.

\begin{figure}
  \includegraphics[width=5.5 in]{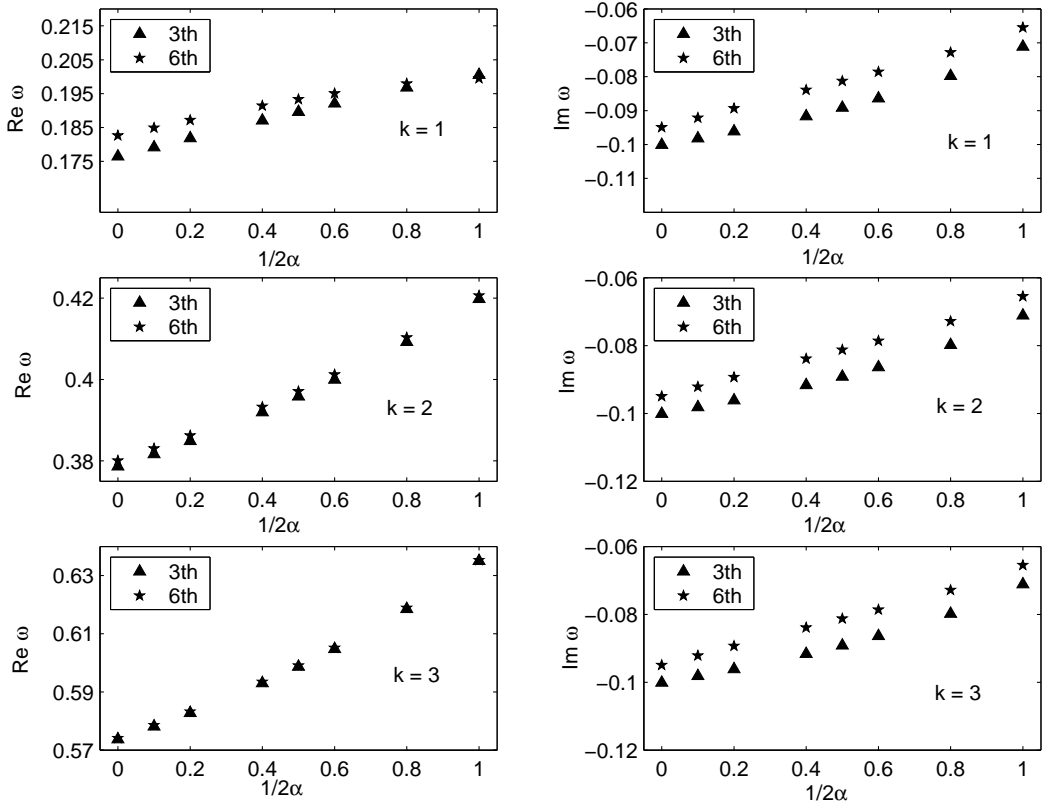}\\
  \caption{The real parts $Re \omega$ and the imaginary parts $Im \omega$ versus radial $r$ under
  the third orders (solid triangle) and the sixth orders (solid pentagram) WKB approximations. }\label{fig4}
\end{figure}

For the changed $\alpha$, we treat $1/2\alpha$ changed in $[0, 1]$ as a whole. Three important kinds low-lying modes: ($k = 1, n= 0$),
($k = 2, n= 0$) and ($k = 3, n = 0$) are listed in Table II, III, and IV, respectively.
According these three tables, we plot the real part
$Re \omega$ and the imaginary part $Im \omega$ of the third and the sixth order results in Fig.\ref{fig4}.
Here, it should notice that the horizontal abscissa denotes the value of $1/2\alpha$. We impose the interpretations on these data and draw conclusions from them.

(1) The real part $Re \omega$ increases as bigger $1/2\alpha$ and the
absolute value of imaginary part $|Im \omega|$ decreases with increasing $1/2\alpha$,
which indicates these QNM could be lived longer.

(2) The gap between the third results and the sixth order ones is visibly displayed in the imaginary part.
The real parts of them basically have the same values, except for $k = 1$ modes.
In general, the average relative magnitudes of the gaps are approximately given as,
\begin{eqnarray}
  \bigg|\frac{^{6th}\text{Im}(\omega) - ^{3th}\text{Im}(\omega)}{^{3th}\text{Im}(\omega)}\bigg| &
  \approx & 10\%, \\
  \bigg|\frac{^{6th}\text{Re}(\omega) - ^{3th}\text{Re}(\omega)}{^{3th}\text{Re}(\omega)} \bigg| &
  \approx & 0.
\end{eqnarray}
Hence, the orders of WKB
approxmations have the tremendous bearing on the damping rate, more than on the actual frequency.

Moreover, when horizontal abscissa of Fig.\ref{fig4} approaches Schwarzschild
case, namely $1/2 \alpha \longrightarrow 0$
(or $\alpha \longrightarrow + \infty$), the real frequencies $Re \omega$ decreases and the
damping rate $|Im \omega|$ increases. In another words, the Horava-Lifshitz gravities have the longer lived and more bigger actual
frequency than that of usual Schwarzschild case.
This specific phenomenon also is observed in the massless
scalar field perturbation \cite{Konoplya2,sbchenqnm}.

\section{conclusion}
In this paper, we have investigated fermions tunneling and perturbation in the IR modified Ho$\check{r}$ava-Lifshitz gravity. We summarize what has
been achieved.

(1) For the fermions Hawking radiation, we consider the symmetrical characteristic of spacetime and adopt the action with
the form of Eq.(\ref{21action}). Through the decomposition of Dirac Eq.(\ref{21diraceuation}),
we can
get the imaginary part of fermions action which could help us obtain the tunneling probability
according to Eq.(\ref{21tunnellingprobability}). Then, on the benefit of $\Gamma = \exp (- \beta \omega)$,
the tunneling temperature could be given as Eq.(\ref{21hawkingtemperture}). So, according to the first
law $d M = T dS$, we have a tunneling entropy Eq.(\ref{211logentropy}) naturally.
It is interesting that the tunneling Hawking temperature and tunneling entropy
are agreement with that obtained by calculating surface gravity.

(2) For the fermions perturbation, we obtain lowly damped
quasinormal modes by using the sixth orders WKB approximations, as well as the third orders formulas.  In order to get a detail analysis on these
obtained quasinormal frequencies, we adopt two kinds of methods: one is to fix the Ho$\check{r}$avaparameter $\alpha$
(fixed $\alpha$) and another is to change $\alpha$ in the range $[0, +\infty]$ (varied $\alpha$).

For the fixed $\alpha$ case,
the results turns out that the impact of the Horava-Lifshitz gravity on quasinormal frequencies
is quite. This is profoundly manifested in the following ways: the actual frequencies becomes bigger and
the damping rate becomes more slower which indicates these lowing modes could be lived longer than
that of usual Schwarzschild. This fact also could be explained by the perturbation potential $V_1$ Eq.(\ref{potential1})
illustrated in Fig.\ref{fig1}, which shows the potential contained Horava-Lifshitz gravity (solid lines)
is lower than that of Schwarzschild (dotted lines).

For the varied $\alpha$ case, we have calculated numerically three kinds of important lowing modes
($k = 1, n = 0$), ($k = 2, n = 0$) and ($k = 3, n = 1$) by through the third and sixth orders
WKB approximations. The result listed in Tables II, III and IV show three facts as follows.
(i) With bigger parameter $1/2\alpha$, the real part of frequencies increases and the absolute
value of imaginary part decreases which is illustrated by Fig.\ref{fig4}. In other words, if
parameter $1/2\alpha$ becomes larger, the actual frequency of QNMs will be larger with more longer
 damping rate.
(ii) The real part is not sensitive to the third or the sixth orders WKB approximates.
This fact also is illustrated in the left sub-plotted curves in Fig.\ref{fig4}, except for some
small $1/2\alpha$ modes of $k = 1$.
(iii) Against the real part, the imaginary part is sensitive
to our WKB approximates methods. The gap of the third and the sixth orders results is unchanged basically.
This fact also is illustrated in the right sub-plotted curves in Fig.\ref{fig4}. In a word, if these specific information could be tested by LIGO, VIRGO, TAMT, GEO600,
it will support Ho$\check{r}$ava-Lifshitz gravity  forcefully and energetically.
\acknowledgments  Project is supported by National Natural Science
Foundation of P.R. China (No.11005088) and Natural Science Foundation of Education Department of Henan Province (No.2011A140022).

\end{document}